\documentclass[fleqn,usenatbib]{mnras}

\usepackage[T1]{fontenc}
\usepackage{xcolor}
\usepackage{graphicx}
\usepackage{subfigure}
\usepackage{amsmath}
\usepackage{newtxtext,newtxmath}

\usepackage{cleveref}
\crefformat{section}{\S#2#1#3} 
\crefformat{subsection}{\S#2#1#3}
\crefformat{subsubsection}{\S#2#1#3}

\newcommand{\msun}{M$_{\odot}$}
\newcommand{\m}{M$_*$}
\newcommand{\ha}{H$\alpha$}
\newcommand{\f}{$JH_{F140}$}
\newcommand{\h}{$H_{F160}$}
\newcommand{\dms}{$\Delta$MS}
\newcommand{\prospector}{\texttt{Prospector}}

\title[sSFR profiles in IllustrisTNG50 vs. 3D-HST]
{Spatially Resolved Star Formation and Inside-out Quenching in the TNG50 Simulation and 3D-HST Observations}

\author[E. J. Nelson et al.]{Erica J. Nelson,$^{1,2}$\thanks{E-mail: erica.june.nelson@colorado.edu}
Sandro Tacchella,$^{2}$
Benedikt Diemer,$^{3}$
Joel Leja,$^{4,5,6}$
Lars Hernquist,$^{2}$
\newauthor
Katherine E. Whitaker,$^{7,8}$
Rainer Weinberger,$^{2}$
Annalisa Pillepich,$^{9}$
Dylan Nelson,$^{10}$
\newauthor
Bryan A. Terrazas,$^{2}$
Rebecca Nevin,$^{2}$
Gabriel B. Brammer,$^{8,11}$
Blakesley Burkhart,$^{12,13}$
\newauthor
Rachel K. Cochrane,$^{2}$
Pieter van Dokkum,$^{14}$
Benjamin D. Johnson,$^{2}$
Federico Marinacci,$^{15}$
\newauthor
Lamiya Mowla,$^{16}$
R\"udiger Pakmor,$^{19}$
Rosalind E. Skelton,$^{18}$
Joshua Speagle,$^{16,18}$
Volker Springel,$^{19}$
\newauthor
Paul Torrey,$^{20}$
Mark Vogelsberger,$^{21}$
Stijn Wuyts$^{22}$
\\
\\
$^{1}$Department for Astrophysical and Planetary Science, University of Colorado, Boulder, CO 80309, USA \\
$^{2}$Center for Astrophysics | Harvard-Smithsonian, Cambridge, MA 02138, USA \\
$^3$Department of Astronomy, University of Maryland, College Park, MD 20742, USA\\
$^4$Department of Astronomy \& Astrophysics, The Pennsylvania State University, University Park, PA 16802, USA\\
$^{5}$Institute for Computational \& Data Sciences, The Pennsylvania State University, University Park, PA, USA\\
$^{6}$Institute for Gravitation and the Cosmos, The Pennsylvania State University, University Park, PA 16802, USA\\
$^{7}$Department of Astronomy, University of Massachusetts, Amherst, MA 01003, USA\\
$^{8}$Cosmic Dawn Center (DAWN), Copenhagen, Denmark\\
$^{9}$Max-Planck-Institut f{\"u}r Astronomie, K{\"o}nigstuhl 17, 69117 Heidelberg, Germany\\ 
$^{10}$Universit\"{a}t Heidelberg, Zentrum f\"{u}r Astronomie, Institut f\"{u}r theoretische Astrophysik, Albert-Ueberle-Str. 2, 69120 Heidelberg, Germany\\ 
$^{11}$Niels Bohr Institute, University of Copenhagen, Jagtvej 128, K\o benhavn N, DK-2200, Denmark\\
$^{12}$Department of Physics and Astronomy, Rutgers University, 136 Frelinghuysen Rd., Piscataway, NJ 08854, USA\\
$^{13}$Center for Computational Astrophysics, Flatiron Institute, 162 5th Ave., New York, NY 10010, USA\\
$^{14}$Astronomy Department, Yale University, New Haven, CT 06511, USA\\
$^{15}$Department of Physics and Astronomy ``Augusto Righi", University of Bologna, via Gobetti 93/2, 40129 Bologna, Italy\\
$^{16}$Dunlap Institute for Astronomy \& Astrophysics, University of Toronto, Toronto, ON M5S 3H4, Canada\\
$^{17}$South African Astronomical Observatory, Cape Town 7935, South Africa\\
$^{18}$Department of Statistical Sciences, University of Toronto, Toronto, ON M5S 3G3, Canada\\
$^{19}$Max-Planck-Institut f\"{u}r Astrophysik, 85740 Garching bei M\"{u}nchen, Germany\\
$^{20}$Department of Astronomy, University of Florida, Gainesville, FL 32611, USA\\
$^{21}$Department of Physics and Kavli Institute for Astrophysics and Space Research, Massachusetts Institute of Technology, Cambridge, MA 02139, USA\\
$^{22}$Department of Physics, University of Bath, Claverton Down, Bath BA2 7AY, UK
}

\date{}
\pubyear{2021}

\begin{document}
\label{firstpage}
\pagerange{\pageref{firstpage}--\pageref{lastpage}}
\maketitle

\vspace{-0.5cm}
\begin{abstract}
We compare the star forming main sequence (SFMS)  -- both integrated and resolved on 1kpc scales --  between the high-resolution TNG50 simulation of IllustrisTNG and observations from the 3D-HST slitless spectroscopic survey at $z\sim1$.  Contrasting integrated star formation rates (SFRs), we find that the slope and  normalization of the star-forming main sequence in TNG50 are quantitatively consistent with values derived by fitting observations from 3D-HST with the \prospector\ Bayesian inference framework. The previous offsets of 0.2-1~dex between observed and simulated main sequence normalizations are resolved when using the updated masses and SFRs from \prospector. The scatter is generically smaller in TNG50 than in 3D-HST for more massive galaxies with \m$>10^{10}$\msun, even after accounting for observational uncertainties.  When comparing resolved star formation, we also find good agreement between TNG50 and 3D-HST: average specific star formation rate (sSFR) radial profiles of galaxies at all masses and radii below, on, and above the SFMS are similar in both normalization and \textit{shape}. Most noteworthy, massive galaxies with \m$>10^{10.5}$\msun, which have fallen below the SFMS due to ongoing quenching, exhibit a clear central SFR suppression, in both TNG50 and 3D-HST. In TNG this inside-out quenching is due to the supermassive black hole (SMBH) feedback model operating at low accretion rates. In contrast, the original Illustris simulation, without this same physical SMBH mechanism, does not reproduce the central SFR profile   suppression seen in data. The observed sSFR profiles provide support for the TNG quenching  mechanism and how it affects gas on kiloparsec scales in the centers of galaxies.
\end{abstract}



\section{Introduction} \label{sect:intro}

Very generally, the fundamental challenge in trying to understand how galaxies form is that it happens over such long timescales. At its present star formation rate, the Milky Way would take over thirty billion years to double its stellar mass \citep[e.g.][]{Licquia:14}. 
No matter the advances in telescope technology, we cannot watch a galaxy through the billions of years of its evolution to see how it builds its bulge and disk, what drives changes in its star formation rate, or how it responds to interactions with other galaxies or changes in accretion rate. 
Various methods have been devised to trace galaxies across cosmic time \citep[e.g.][]{vanDokkum:10a,Leja:13,Behroozi:13,Moster:13,Papovich:15,Wellons:17,Torrey:17b}. But clever as these methods are, they can only tell us about the statistical evolution of a population; they can give us a description of the buildup of a group of similar mass galaxies through time but cannot tell us how it happened. Similarly, the archaeological approach to galaxy evolution tends to be mainly limited to understanding the stellar-mass assembly and chemical evolution of galaxies \citep[e.g.][]{Thomas:99,Graves:09a,Trager:09,Pacifici:16}.

A complementary approach to this problem is to simulate galaxy formation rather than observe it. Simulating a universe in a box allows us to track galaxies through time to see how they grow and determine the key physical processes driving that growth. Cosmological hydrodynamical simulations evolve a box of dark matter, gas, stars, and supermassive black holes through time using gravity and hydrodynamics. Refining these simulations has informed us about the plethora of physical processes involved in galaxy formation. However, it is only in the last decade that hydrodynamical simulations have begun to produce  galaxies with realistic morphologies \citep[e.g.][]{Governato:10,Brooks:11,Guedes:11,Aumer:13a,Christensen:14,Hopkins:14,Vogelsberger:14a,Vogelsberger:14b,Genel:14,Sijacki:15,Schaye:15,Crain:15,Khandai:15,Dave:16,Dubois:16}.  

In general, these simulations come in two types: cosmological volumes focusing on population statistics at the expense of resolution, and zoom-in simulations focusing on individual galaxies at the expense of population statistics. 
With gradual improvements in physical models, computational methods, and spatial resolution, it has become possible to simulate a cosmological volume with resolution sufficient to study the structural evolution of galaxies (thousands of  galaxies at sub-kpc resolution). 
TNG50 is the highest resolution simulation of the IllustrisTNG project, covering a 50 Mpc box with a median spatial resolution of $\sim100$ pc (TNG: \citealt{Weinberger:17,Pillepich:18a,Springel:18,Naiman:18,Marinacci:18,DNelson:18,DNelson:19b}, TNG50: \citealt{Pillepich:19,DNelson:19}).
Studying the structural evolution of galaxies and its relation to the regulation of star formation requires both the spatial resolution and the population statistics afforded by TNG50. However, before it is used for this purpose, the simulation needs to be validated against key observables.

In the space of colour and magnitude, we have long known that galaxies occupy the `blue  cloud' and `red sequence' \citep[e.g.][]{Strateva:01,Kauffmann:03,Blanton:03,Bell:04,Faber:07,Brammer:09,Whitaker:11,Taylor:15}. With improvements in our ability to constrain the physical properties of galaxies, we have found that this blue `cloud' in colour-magnitude space resolves itself into a `sequence' in SFR-M space.  This so-called `star-forming main sequence' (SFMS) is a somewhat sublinear relation between log(SFR) and log(M). The normalization declines with time reflecting slower relative growth rates of galaxies through cosmic time \citep[e.g.][]{Noeske:07,Daddi:07,Salim:07,Rodighiero:11,Karim:11,Wuyts:11b,Whitaker:12,Whitaker:14,Speagle:14,Shivaei:15,Tasca:15,Schreiber:15,Tomczak:16,Lee:18}.

The star-forming main sequence has a scatter of about a factor of two (which has been deemed `tight'). However, not all galaxies reside on the main sequence at all times, they form stars more rapidly or slowly over the course of their assembly history. What drives their evolution through this plane, however, remains uncertain. Star formation across the main sequence has been proposed to be regulated by mergers; episodes of `compaction' and inside-out quenching; bursty star formation; self regulation by accretion and outflows; and variations in dark matter halo formation times \citep[e.g.][]{Hernquist:89,Wuyts:11b,Sparre:15,Sparre:17,Tacchella:16a,Tacchella:20,Nelson:16b,Orr:17,Mathee:19}

Recently we have developed the ability to place spatially resolved constraints on the star forming main sequence. This became possible owing to the capability of mapping tracers of star formation and stellar mass in representative samples of galaxies with e.g. HST/WFC3, VLT/SINFONI, SDSS IV/MaNGA, and  in particular of measuring where star formation happens in galaxies on, above, and below the star forming main sequence at different masses \citep{Nelson:16b,Tacchella:18a,Ellison:18,Belfiore:18,Abdurrouf:18,Morselli:19}.  This tells us where star formation occurs when galaxies are forming stars normally and where it is enhanced and suppressed relative to the existing stars. Galaxy structure and the regulation of star formation appear to be intimately coupled and this measurement provides a direct link between them.  

The integrated and resolved star forming main sequence depends on several key aspects of galaxy formation models: where gas settles in galaxies, feedback, and the conversion of gas into stars. 
For this reason, the star forming main sequence has been used regularly to validate simulations \citep[e.g.][]{Torrey:14,Sparre:15,Schaye:15,Somerville:15,Dave:16,Donnari:19}. However, while the star forming main sequence in recent state-of-the-art simulations has been found to match observations qualitatively, it does not usually match quantitatively, typically having a normalization which is $0.1-1$\,dex too low especially at $z=1-3$ \citep{Somerville:15}. Specifically compared to the chosen observations in each work, it is 0.1-0.5 dex lower in Illustris at $1<z<2$ \citep[][]{Torrey:14,Sparre:15}, 0.2\,dex lower in EAGLE at $0.05<z<0.3$ \citep{Schaye:15}, $0.2-1$\,dex lower in SIMBA \citep{Dave:19}, and $0.2-0.5$\,dex lower in TNG100 \citep{Donnari:19}. 

It is unclear whether this is due to problems with the simulations or uncertainties in the observations. Given the phenomenological nature of prescriptions for AGN and Stellar feedback and star formation, it is entirely possible that this points to a problem with the simulations.  On the other hand, measurements of star formation rates from observations are notoriously difficult and are typically subject to a factor of two systematic uncertainty. The other dimension of the SFR-\m\ plane, stellar mass, is better constrained but still has systematic uncertainties of at least 0.1 dex \citep[e.g.][]{Muzzin:09}.
Resolved measurements of star formation across the main sequence have also been compared between observations and simulations yielding qualitative disagreements. While observations generally find specific star formation rate (sSFR) profiles that are flat or rising on and below the star forming main sequence respectively, simulations typically find they are falling with radius, in particular below the main sequence and in sharp contrast to observations \citep[FIRE, Illustris, SIMBA respectively:][]{Orr:17,Starkenburg:19,Appleby:20}.
In order to use a simulation to understand the structural evolution of galaxies and the regulation of star formation, we must be confident it reproduces the integrated and resolved star forming main sequence. We must understand where the simulation can or cannot reproduce these key observables and determine why in order to physically interpret the observations based on the models we compare them with.

With high quality observational measurements and simulations with improved resolution and prescriptions for feedback, in this paper we compare the integrated and resolved star forming main sequence from the Illustris TNG50 magneto-hydrodynamical cosmological simulation to that inferred from observations as part of the 3D-HST survey at $z\sim1$. We first compare the normalization, slope, and scatter of the integrated star forming main sequence. We then compare the resolved specific star formation rate radial profiles of galaxies below, on, and above the main sequence.

Hubble, Spitzer, and Herschel have spent thousands of hours imaging the CANDELS/3D-HST extragalactic legacy fields to place the best possible photometric constraints on the UV-FIR spectral energy distributions (SEDs) of galaxies which we model to derive physical parameters. This community investment provides the backbone of this work. Two additional features make our work unique. First, owing to the new Bayesian inference framework \prospector, we now have improved measurements of the star formation rates and stellar masses of galaxies changing observed estimates of the star forming main sequence \citep[][]{Johnson:17,Leja:17,Leja:19b,Johnson:20}. Second, owing to the Hubble space telescope WFC3/G141 grism and multiband imaging, we now have spatially resolved measurements of the specific star formation rates for large samples of galaxies across the star forming main sequence \citep[e.g.][]{Nelson:16b}.

The observations on which this comparison is based are from the 3D-HST survey. The 3D-HST survey is a 248 orbit survey with the Hubble Space Telescope (HST) Wide Field Camera 3 (WFC3) grism which provided spatially resolved near-infrared spectra for 200,000 objects in the five major extragalactic legacy fields \citep[][]{Brammer:12a,Skelton:14,Momcheva:15}. At $0.7<z<1.5$ these spectra can be used to create \ha\ emission line maps, which trace where star formation is occurring \citep[e.g.][]{vanDokkum:11,Nelson:12,Nelson:13,Brammer:12b,Lundgren:12,Schmidt:13,Wuyts:13,Vulcani:15b,Vulcani:16b}, for 3200 galaxies with $9<$log(M$_*)<11$ across the star-forming main sequence \citep[e.g.][]{Nelson:16b}, over an order of magnitude more than was previously possible.
Enormous gains were made in our ability to map \ha\ emission with near infrared integral field units on 10-meter class telescopes with adaptive optics \citep[e.g.][]{Forster-Schreiber:06,Forster-Schreiber:09,Forster-Schreiber:11a,Forster-Schreiber:11b,Tacchella:15a}. The information content  in these deep spectra allows detailed study of physical processes in those objects similarly to cosmological zoom simulations. As with the computational cost of zoom simulations, the observational costs of these types of observations are high, limiting the statistics to of order $\sim100$ galaxies. The WFC3/G141 grism provided another window into this problem that is well matched to cosmological simulations like TNG. The slitless spectra provide spatially resolved emission line diagnostics for all objects in its field of view, dramatically increasing the multiplexing capabilities. On a strategic level, we note that with a richer information content, these VLT/SINFONI observations for tens of  galaxies are well-matched to zoom simulations while HST/WFC3 grism observations are well-matched to simulations of cosmological volumes with thousands of galaxies. With a similar resolution and volume, TNG50 and 3D-HST are particularly well-suited to each other. 

This paper is organized as follows. In \S\ref{sect:data}, we describe the data used for this project and how we infer physical properties of galaxies from them. In \S\ref{sect:sims} we describe the TNG50 simulation. In \S\ref{sect:compsfms}, we compare the integrated star forming main sequence slope, normalization, and scatter in TNG50 to observations from 3D-HST/\prospector. In \S\ref{sect:compprof} we compare the specific star formation rate profiles of  galaxies below, on, and above the star forming main sequence between TNG50 and 3D-HST. In \S\ref{sect:summary} we summarize our findings. 

\section{Observational Data} \label{sect:data}

\subsection{Integrated Quantities}\label{subsec:dataint}
In this paper, the key quantities are redshifts, stellar masses, and star formation rates, both integrated and resolved in the case of the latter two.
The 3D-HST+CANDELS dataset is particularly well designed for deriving these quantities in the $z=0.5-2$ Universe as it has 1 kpc spatial resolution imaging and spectroscopy in the rest-frame optical that is key for inferring structural stellar population properties.  
CANDELS is a 902 orbit HST survey providing optical and near-infrared imaging \citep{Grogin:11,Koekemoer:11}. 3D-HST is a 248 orbit HST survey including near-infrared imaging and slitless spectroscopy over the same area \citep{vanDokkum:11,Brammer:12a,Skelton:14,Momcheva:16}. 
These surveys cover five major extragalactic fields AEGIS, COSMOS, GOODS-N, GOODS-S, and UDS which, crucially, have a wealth of publicly available data from the ultraviolet through the infrared \citep[ see Table 3 of Skelton et al. 2014 for additional references]{Giavalisco:04,Whitaker:11,Grogin:11,Koekemoer:11,Brammer:12a,Ashby:13,Skelton:14,Momcheva:16,Oesch:18,Whitaker:19}. 

Redshifts are derived from template fits to the combination of photometry and near infrared slitless spectroscopy \citep{Momcheva:16}.  
Galaxy stellar masses and star formation rates are derived by modeling the 0.3--24$\mu$m (UV-IR) spectral energy distribution (SED) from the observed photometry. Aperture photometry was performed on PSF-matched images to measure consistent colours across passbands. For the HST imaging, a $0\farcs7$ diameter aperture was used and an aperture correction was performed to arrive at the total flux \cite[see][for many more details]{Skelton:14}. To determine stellar population parameters, the SED is fit using with the Bayesian inference framework \prospector\ \citep{Johnson:17} as presented in \citep{Leja:19b}. \prospector\ uses the Flexible Stellar Population Synthesis code \citep[FSPS]{Conroy:09} to construct a physical model and the nested sampler \texttt{dynesty} to sample the posterior space \citep{Speagle:20}.
This model includes a non-parametric star formation history, a two-component dust attenuation model with a flexible attenuation curve, variable stellar metallicity, and dust emission powered by energy balance \citep[see ][for more details]{Leja:17}. With this new model, our new catalogs have systematically higher stellar masses and lower star formation rates than previous versions \citep{Leja:19b}. In this work we use the SFRs averaged over the last 30 Myr. 

\subsection{Mapping stellar mass and star formation}
In this paper we compare specific star formation rate profiles of galaxies across the star forming main sequence from TNG50 to observations at $z\sim1$. Deriving specific star formation profiles observationally is challenging due primarily to the difficulty of mapping star formation. 
Our process for deriving sSFR profiles for this comparison builds on \cite{Nelson:16b}, so we refer the reader there for details. The primary update is that we use spatially resolved SED fitting to derive stellar mass maps and perform a dust correction to the \ha\ emission to map star formation. We summarize our methodological choices and their impact below and briefly describe the rest of the analysis and the data from whence it came, with an emphasis on what is new and what is certain or uncertain.

Our aspiration here is to compare sSFR profiles from TNG50 to the real Universe, meaning that we need to map stellar mass and SFRs from observations. We map stellar mass and star formation in two ways, with one method closer to the data and the other with more layers of interpretation. 
In both of these analysis tracks we stack maps, correct for the effects of the point spread function (PSF) on the stack, and then construct radial surface brightness profiles. In the following section, we first describe the different ways we map sSFR and then describe the stacking, PSF-correcting, and profile extraction.

\subsubsection{Resolved sSFR from maps of \ha\ equivalent width}\label{subsec:hamap}

The method closest to the data is to simply use maps of \ha\ equivalent width as a proxy for sSFR. 
Hot young stars photoionize their surrounding gas. The recombination and subsequent cascade of electrons in hydrogen atoms produces the \ha[$6563$\AA] emission line (amongst others) which is thus a tracer of stars formed in the past $\sim10$ million years. At the same wavelength, the rest-frame R-band continuum, light from from the longer-lived, lower-mass stars that make up the bulk of the stellar mass become more important, making it an oft-used tracer of the distribution of stellar mass \citep[e.g.][]{vanderWel:14}. Here we trace this redshifted R-band emission with the WFC3/$JH_{F140}$ filter. The quotient of these, \ha/$JH_{F140}$, which we will here call the \ha\ equivalent width (EW(\ha)) hence traces sSFR. 

The key innovation here is the ability to map the \ha\ emission line, a tracer of star formation, in large samples of galaxies. We do this using the slitless spectroscopy from the 3D-HST survey which provides spatially resolved maps of emission lines for everything in its field of view \citep[e.g.][]{Nelson:12,Nelson:13,Nelson:16b,Brammer:12b,Lundgren:12,Schmidt:13,Wuyts:13,Vulcani:15b,Vulcani:16b}. Due to its large multiplexing capacity and unbiased sampling, this mode has grown increasingly popular on HST and likely will on JWST as well. 
The grism (a portmanteau of ``grating" and ``prism"), is a spectral element in the WFC3 IR channel filter wheel dispersing incident light onto the WFC3 detector, and as such providing spectra for all objects in the field of view. 
This observing mode features a unique combination of HST's high native spatial resolution and the grism's low spectral resolution: $\sim$1 kpc and $\sim$1000 km/s at $z=1$, our redshift of interest. This means that for all galaxies in our sample we will get a map of the spatial distribution of line-emitting gas. Because of the  low spectral resolution, these spectra contain virtually no kinematic information; besides e.g. >1000 km/s outflows, the entire velocity structure of the galaxy will be contained in a single spectral resolution element. Hence we obtain maps of the emission lines of all galaxies in the field of view which are redshifted into the wavelength coverage of the grism. 

The wavelength coverage of the G141 grism ($1.15-1.65\, \mu$m) samples redshifted \ha\ at $0.7<z<1.5$.
The spectra of all objects in the field are forward-modeled based on imaging. This provides the extraction window for each spectrum based on the geometric transformation onto the detector. Furthermore, because there is nothing blocking the light from other objects, many of the spectra overlap or ``contaminate" one another. The forward-modeling also maps where contaminating flux from other objects will fall on the 2D spectrum of the object of interest. All pixels predicted to have contaminating flux more than a third of the background are masked. Finally, the continuum light of a galaxy is modeled by convolving the best-fit SED without emission lines with its HST image at the same wavelength (combined $J_{F125W}/JH_{F140W}/H_{F160W}$). We subtract the continuum model from the 2D grism spectrum which simultaneously removes the continuum emission and corrects the \ha\ maps for underlying stellar absorption. What remains for all 3200 galaxies at $0.7<z<1.5$ is a map of their \ha\ emission. One complication of the low spectral resolution is that N{\sc ii} and \ha\ are blended and S{\sc ii} and \ha\ are separated by three resolution elements. To mitigate this,  we use a double wedge mask along the dispersion direction covering S{\sc ii}. The overall contribution of N{\sc ii} has less of an impact because the total map is scaled to the integrated SFR measured from \prospector and the mask decreases the impact of very high ratios extending emission in the dispersion direction. Radial gradients in N{\sc ii}/\ha, on the other hand do matter. We account for these in \S\ref{subsec:dataprof}. 

More details on the reduction and analysis of the 3D-HST grism spectroscopy are available in \cite{Brammer:12a,Momcheva:16}; more details on the creation of \ha\ maps are in \cite{Nelson:16b}.
Mapping the \f\ emission is much more straightforward. Stamps are  cut around the objects in the interlaced frames. Light from nearby objects is masked according to the SExtractor segmentation map. 

This first method for mapping sSFR comes straight from the data: it is simply the quotient of the measured \ha\ map and the measured \f\ map. No dust correction is done to either the \ha\ or the \f\ maps, with the assumption that they are subject to similar dust attenuation because they are at the same wavelength (modulo differential extinction toward H{\sc ii} regions) hence the dust attenuation multiplier cancels out in the quotient. 

\subsubsection{Resolved sSFR from spatially resolved SED fitting} 

The effect of dust attenuation in principle cancels when scaling the observed \ha/\f\ directly to sSFR (as described in the previous section). However, in addition to dust, the continuum light, which we are scaling to stellar mass, is also subject to age gradients which affect the mass-to-light ratio ($M/L$). In particular, the centers of galaxies are typically observed to be older than their outskirts \citep[e.g.][]{Wuyts:12,Cibinel:15,Tacchella:15b}. Older stars have a higher $M/L$ meaning that the stellar mass is more concentrated than the light. Consequently the actual sSFR profiles could be more centrally depressed than the observed profiles of \ha/\f.  

Our second method attempts to mitigate the effects of dust and stellar age on the observed light using spatially resolved spectral energy distribution (SED) modelling to map the stellar mass and dust attenuation in our galaxies. Spatially resolved SED modelling is done using the eight band HST imaging described in \S\ref{subsec:dataint}. 
This methodology is described in detail in \cite{Cibinel:15}, but we outline it here for completeness. 
Image postage stamps are cut from the mosaics in each HST band convolved by PSF-matching to the resolution of the reddest band, \h, which has the lowest resolution. 
The images are adaptively smoothed using Adaptsmooth \citep{Zibetti:09} requiring $S/N>5$ in each spatial bin in the $H_{F160W}$ image, which has the highest $S/N$. The SPS code LePhare \citep{Arnouts:99,Ilbert:06}
is run on the photometry in each spatial bin using the \cite{Bruzual:03} synthetic spectral library, a \cite{Chabrier:03} initial mass function, a \cite{Calzetti:00} dust law and three metallicity values ($Z=0.2,0.4,1\,Z_\odot$). The star formation history is parameterized as a delayed exponential $(t/\tau^2)\,{\rm exp}\,(-t/\tau)$ having a characteristic timescale $\tau$  with 22 values between 0.01 and 10 Gyr and a minimum age of 100\,Myr. 

We use the model $E(B-V)$ maps to correct our \ha\ maps for the effects of dust using 
$$ A_{cont} = k(\lambda)E(B-V) $$
$$ A_{extra} = 0.9A_{cont}-0.15A_{cont}^2$$
$$ F(H\alpha)_{\rm intr} = F(H\alpha)_{\rm obs} \times 10^{0.4A_{cont}} \times 10^{0.4A_{extra}}$$

\noindent where $A_{cont}$ is the dust attenuation toward  the stellar continuum at the wavelength of \ha. $k(\lambda)$ is computed using the \cite{Calzetti:00} dust  attenuation law $k(H\alpha = 6563$\AA$)=3.32$.
$A_{extra}$ is the amount of extra attenuation towards H{\sc ii} regions calculated  using \cite{Wuyts:13}.
These dust corrected maps of SFR(\ha) are then divided by then the SED-modeled stellar mass to get the sSFR. Both are scaled to the integrated values from \prospector.

\subsection{Selection}

We select galaxies in the redshift range $0.75<z<1.5$ for which we can map the \ha\ emission line using the HST/G141 grism. We confine this analysis to the mass range $9<$log(\m)$<11$; the lower boundary is driven by our completeness limit \citep{Tal:14}, the upper by number statistics. Here and for the remainder of this paper when we refer to log(\m), it is in units of \msun. 
Here we are interested in an analysis of the SFMS rather than the star formation properties of the full population of galaxies, so we select only those  galaxies  which  are  actively forming stars. 
We do this according to a doubling time criteria, specifically by comparing the doubling time to the age of the Universe \citep{Tacchella:19}. We use a slightly less restrictive criteria to encompass the tail of the distribution to low SFRs:
$$ t_{\rm double} < 20 t_{\rm Hubble}(z) $$
This corresponds to a galaxy's current star formation rate doubling its mass in 20 Hubble times (or adding 5\% to its mass in a Hubble time). 
This is the extent of the selection criteria applied for \S\ref{sect:compsfms} comparing the integrated SFMS in observations and TNG50. 

For \S\ref{sect:compprof} comparing sSFR profiles across the main sequence, a few additional selection criteria are required on the observational side. We remove all galaxies flagged as having unreliable photometry as well as galaxies with with X-ray luminosity $L_x>10^{42.5}{\rm erg\,\,s}^{-1}$ or \ha\, emission line widths of $\sigma>2000$\,km/s likely indicating that emission from an active galactic nucleus (AGN) will contaminate the central \ha\ flux we interpret as star formation. 
For the \ha\ maps, we also reject galaxies whose spectra are too badly contaminated (See \S\ref{subsec:hamap}). Together these  criteria result in a selection of $\sim3200$ galaxies. Finally we note that we have maps of $E(B-V)$ in only two of our five fields, GOODS-N and GOODS-S, where there the HDUV program provides UV data. 

\subsection{Stacking \& specific star formation rate profiles}\label{subsec:dataprof}

We stack galaxies across the main sequence in bins of stellar mass and position with respect to the SFMS (\dms).  Stellar mass bins are 0.5 dex from log(M/M$_{\odot}$)=9-11. We fit the SFMS as described in \S\ref{sect:compsfms} and divide the galaxies into bins below, on, and above the main sequence according to log(\dms) [-0.8,-0.4], [-0.4,0.4], and [0.4,1.2], respectively. 
To create each stack, we take a pixel by pixel mean of all maps in that bin.  Many pixels in a given map are masked so we also make a mean stack of the masks and divide this out to correctly normalize the mean in each pixel. 
No weighting is done for ease of comparison to the simulations. 

A key step in this process is to correct observations for the effect of the point spread function (PSF). The PSF blurs images, resulting in dense regions appearing less dense (and vice versa of course). Our method for correcting for the effect of the PSF uses a parametric model to account for the effects of the PSF on the radial light distribution. To do this, we fit the light distribution (or derived physical quantity) of each stack with a S\'ersic model \citep{Sersic:68} using \texttt{galfit} \citep{Peng:02}. We fit a single S\'ersic model letting the brightness, effective radius, S\'ersic index, centroid, projected axis ratio, and position angle be free and forcing the background level to be zero. The fit is found by convolving each model with the PSF and computing reduced $\chi^2$. All images are background subtracted and their backgrounds have been tested and found to be zero. Forcing the background to zero allows \texttt{galfit} less freedom to fit the wings of the profile. With these best fit parameters, we create a model not convolved with the PSF and add the residuals from the fit. This means that regions of the fit in which the image deviates from the model will be accounted for. The resulting ``PSF-corrected'' image will have the bulk of its light corrected for the PSF but the residuals will not be \citep[e.g.][]{Szomoru:10}.

There are of course several shortcomings with this methodology. First, this method corrects based on a single, axisymmetric S\'ersic profile. This is a reasonable approximation for the mass profile of a high redshift galaxy but real galaxies are of course more complicated. This model effectively reconstructs the radial profile of the light but will not e.g. deconvolve non-axisymmetric features at larger radii, like clumps or spiral arms. While this means that individual images are not as they would be without the PSF, we average over these types of features twice: once in the stack and again in computing the radial profile, so it is not important for this analysis. 
Because on average the profiles of both mass and star formation peak in the centers of galaxies \citep[see e.g.][]{Nelson:16b}, the thing that is most important for us is to replace the light into the center so it is less important for e.g. spiral arms or clumps at large radii to be deconvolved from the PSF. 
Second, because we stack galaxies with different radii, the stack will have a steeper profile than any of the galaxies do intrinsically. The resulting fit will typically have larger S\'ersic index than the average of individual  galaxy images and plausibly put too much light back in the center. 
We acknowledge that this step may induce a feeling of unease in the uninitiated but it is necessary and the best we can do with current tools. 
Ideally, our SED modelling would account for the effects of the PSF so this step would not be required, however a tool of this kind does not yet exist. 

Radial profiles are computed in circular apertures.
To generate specific star formation rate profiles from the equivalent width profiles, we scale the integral of the \ha\ profile to the mean total star formation rate from \prospector\ described above and the $JH_{F140}$ to the stellar mass. 
We also normalize the SED modeled profiles of stellar mass and star formation to the mean integrated values from \prospector.
Error bars are computed by bootstrap resampling the stacks.
The sSFR profiles are the SFR profiles divided by the \m profiles. 

To summarize (and make the order of operations clear): we make maps of \ha, F140W, stellar mass, and dust attenuation for all galaxies where they are available and then stack all available maps for a given bin. For method two, the stacked dust attenuation map is applied to the stacked \ha\ map. Next, all stacks are PSF-corrected, the radial profiles are computed for each \ha, \f, stellar mass, and \ha\ corrected for dust, and finally quotients are performed for each pair.

Two observational issues merit a (somewhat) brief discussion before moving on, with both most strongly affecting the sSFR profiles of massive galaxies. First, it is possible that some fraction of the central light comes from an AGN: both the broad band emission and in particular the \ha\ emission. To estimate the possible extent of this effect, we subtract observational estimates of the contribution of AGN to our observed \ha\ emission from the literature. \cite{Forster-Schreiber:14} and \cite{Genzel:14b} find that in their sample of  $z\sim2$ galaxies with a detected broad velocity component, an average of 37\% of the nuclear \ha\ flux comes from this broad component that they attribute to AGN-driven winds (rather than star formation). Additionally, because of the low resolution of the G141 grism, the \ha\ line we observe is contaminated by N{\sc ii}. In these same studies, the authors find nuclear N{\sc ii}/\ha\ = 0.55 in stacks of galaxies with a broad line detection. That being said, in the galaxy population writ large, \cite{Forster-Schreiber:19} find fairly flat N{\sc ii}/\ha\ gradients. \cite{Genzel:14b} find 35\% of galaxies with 10.5<log(M$_*$/M$_{\odot}$)<11 have a broad component. Accounting for these effects reduces the observed central sSFR by 25\%.  We use this as the default in our analysis but note it has a minimal effect. 

Second, age gradients will affect the observed sSFR profiles in high mass galaxies. Because older stellar populations emit less light per unit mass and we expect the centers of massive galaxies to be older than their outskirts, there is likely more stellar mass in the centers of these galaxies than we infer from the \f\ light profiles. This can be seen in the sSFR profiles based on resolved SED fitting that get more centrally depressed at high mass on and below the main sequence. Above the main sequence, on the other hand, central dust obscuration becomes an issue at high masses. As can be seen in Fig.~\ref{fig:profs_comp}, the sSFR profile based on SED modelling has a higher central sSFR than that based on EW(\ha). Because of the importance of these effects at  high masses, we take the SED modeled sSFR profiles as the  default for Fig.~\ref{fig:ssfrprofs_highmass}.


\section{Simulation data} \label{sect:sims}

\subsection{The TNG50 simulation} \label{subsec:tng50}

TNG50 is a magnetohydrodynamical cosmological simulation of galaxy formation. It is the highest resolution member of the IllustrisTNG family (TNG from now on), The Next Generation of the Illustris project . Henceforth the original Illustris simulation will be referred to as simply Illustris \citep{Vogelsberger:14a,Genel:14,Sijacki:15}. The TNG model was built on the successes of the original Illustris model. However, a few issues with the star formation rates and structures of galaxies in the original Illustris simulation were soon noticed in comparison to observations: the effective radii (of the stellar mass) in Illustris were larger than observed \citep{Pillepich:18a,Genel:18}, the distribution of galaxy colors showed only a weak bimodality between red and blue \citep{Vogelsberger:14b, DNelson:18}, and the normalization of the star-forming main sequence was too low at $z=1-2$ \citep{Sparre:15} in comparison to observational estimates. Thus for TNG, the models for feedback from star formation and AGN were modified as described below. Furthermore, the parameters of the TNG model were chosen to provide a better match to a few key observations including the cosmic star formation history and the following at $z=0$: the galaxy stellar mass function, stellar mass -- halo mass relation, supermassive black hole -- galaxy mass relation, size -- stellar mass relation, and gas fraction within group-mass halos.

TNG50 evolves dark matter, gas, stars, black holes and magnetic fields from $z=127$ to 0. 
With a cubic volume of 51.7 Mpc side length, and a density-dependent resolution in galaxy star forming regions of $70-140$ pc, TNG50 provides resolution typical of zoom simulations of single galaxies for 1600 galaxies with $10^9<{\rm m}<10^{10}$ \msun~and 530 with $10^{10}< {\rm  m}<10^{11}$ \msun~at $z\sim1$.
The baryon mass resolution is $8.5\times10^4$\msun, the gravitational softening length of the dark matter and stars is 0.3 kpc.  This is the most computationally demanding run of the simulation suite, requiring 130 million CPU hours \cite[see][for more details]{Pillepich:19, DNelson:19}. TNG50 evolves a total of 2x2160$^3$ total initial resolution elements, half dark matter particles, half gas cells. They are evolved using AREPO, a massively parallel simulation code optimized for large runs on distributed memory machines \citep{Springel:10}. 

The TNG physical model for galaxy formation includes several physical process thought to be important to galaxy evolution that are implemented at the spatial and mass resolution of the simulation. In addition to gravity and hydrodynamics, the model includes gas cooling and heating, star formation, aging of single age star particles, chemical enrichment of the interstellar medium, and feedback from supernovae and super massive black holes (SMBHs). 
Star formation is modeled with the very simple density threshold-based parameterization of \cite{Springel:03}. In such a prescription, gas is stochastically converted into star particles once its density exceeds $n_H  = 0.1$cm$^{-3}$ on a timescale determined such that the galaxy-wide empirical Kennicutt-Schmidt relation \citep{Kennicutt:89} is broadly reproduced.  

As in any model of galaxy formation, feedback from stars and black holes is essential. Supernova feedback associated with star formation drives galactic scale outflows. In TNG, these outflows are launched directly from star forming gas with energy proportional to the local and instantaneous star formation rate. There are several changes to the Illustris star formation-driven wind model in TNG: the wind injection is isotropic rather than bipolar; the velocity of wind particles now scales with redshift and has a floor; and the energy now depends on metallicity and has a thermal component \citep{Pillepich:18a}. The net result is that the star-formation driven winds in TNG are faster at all masses and times and generally more effective at preventing star formation.

The TNG50 model for feedback from SMBHs is described in detail in \cite{Weinberger:17}: SMBH feedback comes in two flavors, decided by the rate at which the black hole is accreting nearby gas. In the high accretion rate flavor, thermal energy is injected continuously into the surrounding gas, as in Illustris \citep{Springel:05, DiMatteo:05}. At low accretion rates, kinetic energy is injected into the surrounding gas as a time-pulsed, oriented wind in a different random direction at each SMBH timestep. By contrast, in Illustris, highly bursty thermal energy was injected into large ($\sim50-100$kpc) bubbles displaced away from the central galaxy \citep{Sijacki:07}. The new AGN feedback model, particularly the kinetic mode, effectively quenches galaxies that reside in intermediate to high mass halos, including realistic gas fractions \citep{Weinberger:17, Pillepich:18a}.

\subsection{SFRs from TNG50 and other galaxy properties} \label{subsec:props}

In this work, for all galaxies we take the galaxy stellar mass to be the total mass of all star particles that are gravitationally bound to each subhalo, according to the {\sc subfind} halo finder \citep{Springel:01}.
We take the star formation rate to be the sum of the individual star formation rates of all individual gas cells in each subhalo. These are thus instantaneous star formation rates and total masses. While this is what we attempt to measure in observations, as explored in depth in \cite{Donnari:19} and  \cite{Donnari:21}, aperture corrections and imperfect star formation tracers make this inexact, complicating comparisons between observations and simulations. However, we attempt to make our comparison as consistent as possible. 

As with for the 3D-HST data,  we also exclude from the simulated galaxies analysis those with very low SFRs: $t_{\rm double} < 20 t_{\rm Hubble}(z)$. 
This cut in the simulated sample automatically removes completely quenched objects or galaxies whose SFRs are so low that they fall below the resolution limit of TNG50; i.e. objects whose SFR$\equiv 0$. 

Furthermore, when comparing the distribution of SFRs about the main sequence in observations and simulations (\S~\ref{subsec:sfmswidth}) it is essential to account for observational uncertainties. To do this, in observations instead of looking at simply the best-fit value of the SFR, we use the full information about the probability density function (PDF) of the fit. To measure the scatter of the main sequence we sum the probability density function of each galaxy's SFR instead of just looking at the distribution of the best fit values. We apply the same treatment to the SFRs from the simulation. We assign an observed PDF to each SFR and sum the PDFs to determine the scatter of the main sequence in TNG50. In this way we account for observational uncertainties in the comparison between observations and simulations.

\subsection{Radial profiles of sSFR in TNG50} \label{subsec:tngprofs}

The standard approach to making radial profile from simulations is to rotate galaxies to face on then extract profiles in circular annuli. This is of course not how observations are done; observers unfortunately cannot travel out to distant galaxies and rotate them. 
In observations, the light from galaxies as they are oriented on the sky is what falls on our detectors. The blurring done by the point spread function (PSF) will happen on the randomly oriented image in the plane of the detector. For high S/N images, it is possible to do a PSF correction on an individual galaxy image and then deproject it before stacking. With our shallow \ha\ images, however, a PSF correction is not possible on individual galaxy images, it is only possible on a stack. We therefore cannot deproject the observed \ha\ images and instead project the TNG50 particle distributions to mimic the observations.

Maps of stellar mass and star-forming gas cells are made by projecting particles and cells onto a grid of $121^2$ pixels representing a physical size of 60$^2$ kpc, or 0.5 kpc/pixel using the methods developed in \citet{Diemer:17,Diemer:18,Diemer:19} and \citet{Tacchella:19}.
Each particle/cell is distributed onto pixels according to the kernel smoothing used by the simulation. This includes all particles/cells bound to the galaxy according to the {\sc subfind} halo finder. The centroid is defined as the co-moving center of mass of the subhalo calculated by summing the mass weighted relative coordinates of  particles of all types in the subhalo. We project galaxies in the xy plane in the simulation box to mimic the random projection of galaxies in observations. These maps are then mean-stacked and we compute profiles in radial bins. As for the observations, error bars are determined by bootstrap resampling the stacks. We include the three full snapshots in the redshift range of the observations ($z=0.7,1.0,1.5$).

In Fig. \ref{fig:ssfrprofs_orients}, we show the difference between the sSFR profiles derived from the standard face-on projection, the edge-on projection, and the random xy, xz, yz projections. The differences are fairly small but we include this correction for completeness. In particular, this correction has the largest effect in the highest mass bin below the main sequence, which, as we will soon see is a particularly important regime to treat accurately for this comparison.

\section{The Integrated Star Forming Main Sequence: TNG50 vs. 3D-HST}\label{sect:compsfms}

\begin{figure}
    \centering
    \includegraphics[width=0.45\textwidth]{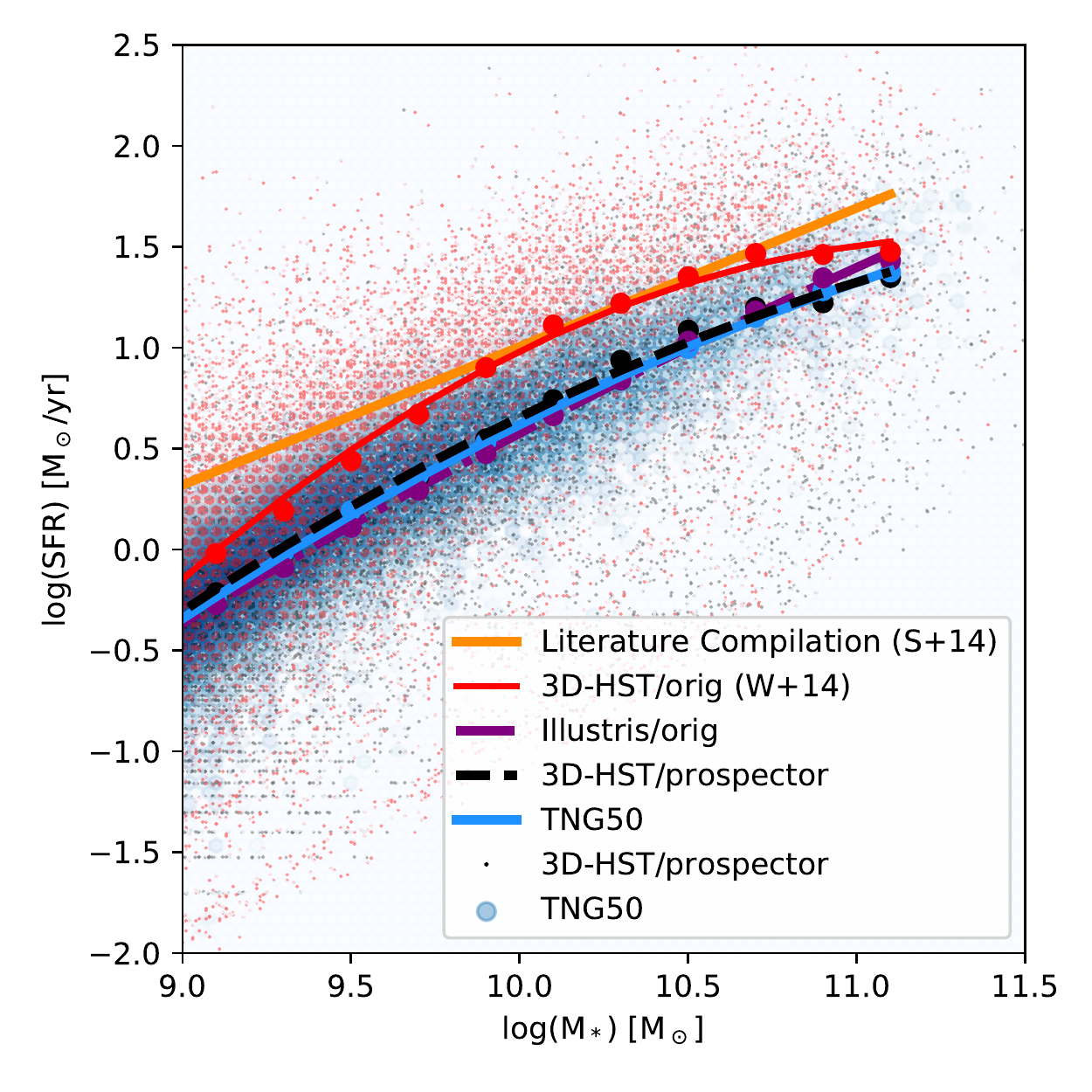}
    \caption{The star forming main sequence (SFMS) in TNG50 (blue points) versus the 3D-HST survey black points at $0.7<z<1.5$. The curves show quadratic fits to the running median star formation rates. For the data we include the original 3D-HST stellar masses and star formation rates \citep[red points and red line][]{Whitaker:14,Skelton:14} and a literature compilation from \citet{Speagle:14}. The new fits from \prospector\ infer stellar masses $0.1-0.3$ dex higher and star formation rates $0.1-1$ dex lower resulting in a star forming main sequence with a normalization systematically lower by $\sim0.2-0.5$ dex. We also show the star forming main sequence from the original Illustris simulation (purple line). The slope and normalization of the SFMS in the simulations is remarkably consistent with observations at this redshift, due to the newly inferred values from the data.}
    \label{fig:sfms_comp}
\end{figure}

Here we investigate similarities and differences in the distribution of galaxies in the star formation rate -- stellar mass plane at $0.7<z<1.5$ between observations from the 3D-HST survey (see \cref{subsec:dataint}) versus TNG50 cosmological hydrodynamical simulations (see \cref{sect:sims}). 
To make this comparison as informative as possible, we analyze the simulations and observations in the same way. 

\subsection{Locus of the star forming main sequence}

We compute running median star formation rates as a function of stellar mass for both samples. 
To define the main sequence, we fit these running medians with a quadratic:
$$ \log({\rm SFR}) = a + b \log(M_*) + c \log(M_*)^2 $$
As described in \S\ref{sect:sims} and \S\ref{sect:data}, in both observed and simulated sample galaxies with very low SFRs are removed from the analysis.

Figure \ref{fig:sfms_comp} shows the distribution of galaxies in the SFR-M$_*$ plane from 3D-HST (dark orange points) and TNG50 (blue points) as well as the SFMS fits to the median SFRs (dark orange vs. light orange curves). 
The median fits are remarkably similar between TNG50 and 3D-HST: they are within 0.1 dex at all masses $9<\log(M_*/M_{\odot})<11$. The SFMS are so similar in fact that one might be tempted to conclude the first author bungled the plotting and used the same array twice. We assure the reader this is not the case: these are truly nearly identical. That being said, there remains of order $\sim0.1$\,dex uncertainty in this comparison due to aperture effects and the timescale on which the SFR is measured, as described in \cite{Donnari:19}.

Let us not lose sight of the main point, however: the SFMS in the TNG50 simulation and observations from 3D-HST are in remarkable agreement. This is surprising given the longstanding $0.1-1$\,dex offset between the SFMSs in observations and simulations at $z=1-2$ \citep[e.g][]{Torrey:14,Sparre:15,Somerville:15,Dave:16,Donnari:19}. So what changed? Let us first  consider the simulations. Illustris and TNG50 main sequences are shown in Fig. \ref{fig:sfms_comp}: light orange vs. yellow curves. There is little change going from Illustris to TNG50 at $z\sim1$; the slope and normalization of the main sequence have remained similar. Turning to the observations, the star forming main sequence from the original 3D-HST catalogs \citep[v4.1.5;][]{Whitaker:14,Skelton:14} as well as a literature compilation \citep{Speagle:14} are also shown. The normalization of the main sequence  at  $z\sim1$ has decreased by $0.2-0.5$\,dex when using the \prospector\ Bayesian inference framework to determine stellar population parameters compared to previous determinations. The offset is minimized when adopting a non-parametric star formation history in the Bayesian inference framework, coupled with accounting for infrared emission due to dust heated by older stellar populations and supermassive black holes rather than star formation \citep[see][for more information]{Leja:19b}. Thus the long-standing $0.1-1$\,dex offset between the SFMS in observations and simulations at $z\sim1$ disappears in this work not due to changes in the simulations but rather to changes in the stellar population parameters inferred from observations. Values for the coefficients in the $z\sim1$ main sequence fit (equation above) are listed in Table \ref{tab:mscoeffs}. 

As shown in \citet{Torrey:14}, the star forming main sequence in simulations is fairly insensitive to the nature of the feedback prescription. The integrated main sequence is thus not a particularly discerning validation of a simulation's feedback model. As we will show in the next section, this  is  not  the case when looking at the resolved properties of  star formation across the  main sequence. Furthermore, \citet{Leja:15} showed that earlier measurements of the star forming main sequence and the evolution of the  stellar mass function were not self consistent in observations: the SFMS dramatically  overpredicted galaxy stellar mass growth. In the simulations they are obviously self-consistent and hence unsurprising that they could not  simultaneously match both the observed main sequence and mass function. With data that are self-consistent, the simulations can match both as they are directly coupled. 

\begin{table}
	\centering
	\caption{Coefficients in the fit to the star forming main sequence in the TNG50 simulation versus observations from the 3D-HST survey both original v4.1.5 and updated with Prospector. $\log{(SFR)} = a + b\log{(M_*)} + c\log{(M_*)}^2$}
	\label{tab:mscoeffs}
	\begin{tabular}{lccr} 
		\hline
		data/sim & a & b & c\\
		\hline
		3D-HST/Prospector & -22.13 & 3.74 & -0.146\\
		TNG50 & -20.46 & 3.38 & -0.127\\
		3D-HST/orig & -37.48 & 6.87 & -0.302\\
		Illustris/orig & -14.21 & 2.08 & -0.060\\
		\hline
	\end{tabular}
\end{table}

\subsection{Width and outliers of the star forming main sequence}
\label{subsec:sfmswidth}

\begin{figure*}
    \centering
    \includegraphics[width=0.915\textwidth,trim = 0 1.5cm 0 0, clip]{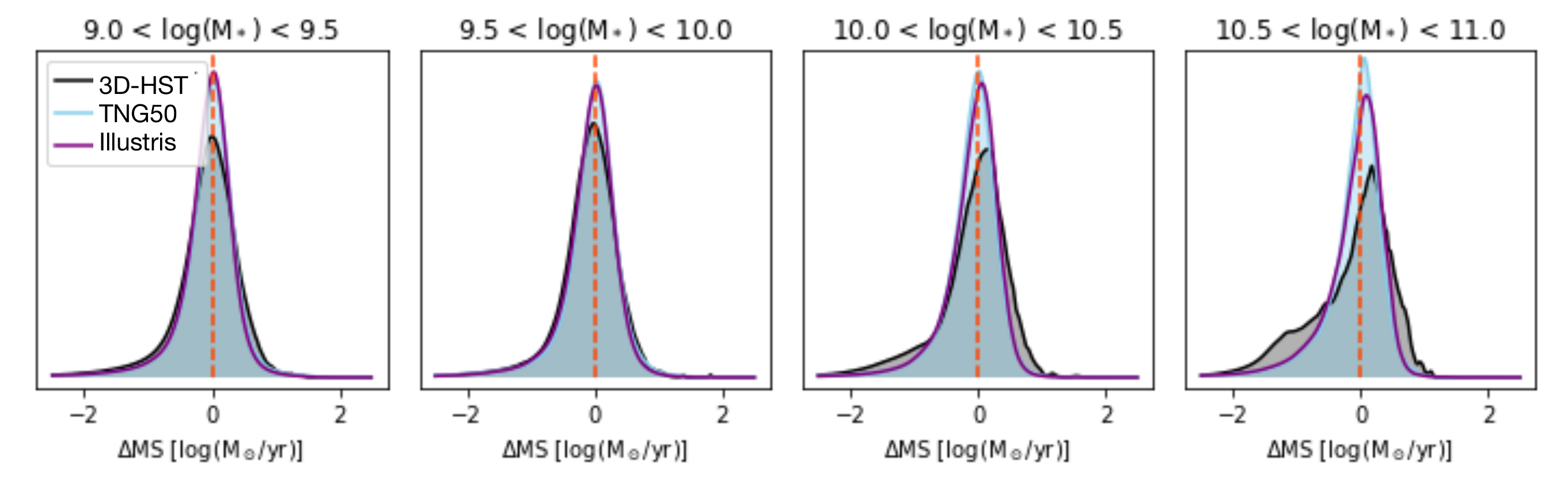}
    \includegraphics[width=0.9\textwidth,trim = 0 0 0 1cm,clip]{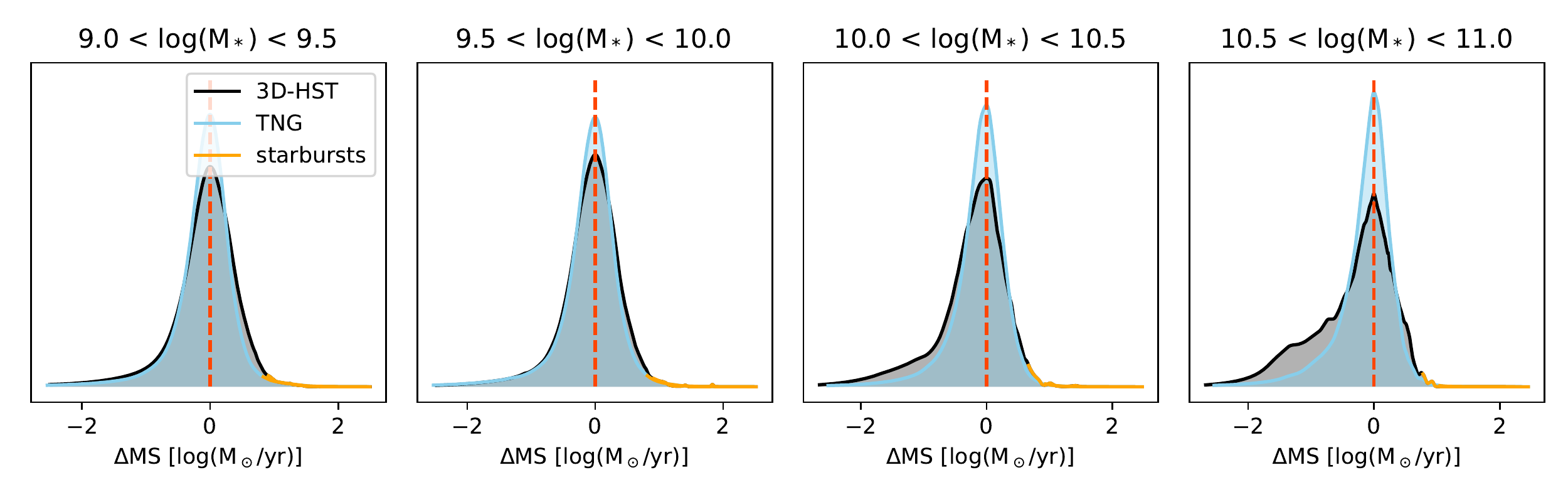}
    \caption{\textbf{Top:} Distribution of simulated and observed galaxies around the main sequence in bins of stellar mass. We contrast the 3D-HST data (black), TNG50 simulation (light blue), and original Illustris simulation (purple). Although the width of these distributions are broadly consistent between the simulations and observations at lower galaxy stellar mass, the simulated scatter is smaller than observed at high ($M_\star > 10^{10}$ M$_{\odot}$). \textbf{Bottom:} As above, except with the distribution shifted to the ridgeline of the distribution of star formation rather than the median. With this shift applied, the shape of the distribution of galaxies above the main sequence is similar between observations from 3D-HST/prospector and TNG50. The orange solid line shows the definition of ``starbursts" used in \S \ref{sect:compsfms} following \citet{Rodighiero:11} and \citet{Sparre:15}: $>2.5\sigma$ above the main sequence.}
    \label{fig:delmsdist}
\end{figure*}

Although the medians are nearly identical between TNG50 and \prospector/3D-HST, the distribution of galaxies about these medians is not, even when accounting for observational uncertainties in our  treatment of the simulations. We look at the distribution of the distance of galaxies from the median fit (\dms) in bins of stellar mass in this Section: see Fig.~\ref{fig:delmsdist}. However, investigating the shape of this distribution requires properly accounting for observational uncertainties. \prospector\  returns a probability density function (PDF) for each parameter it fits. In each bin, we sum the PDFs of SFR normalized to the main sequence fit then normalize the overall distribution to have an area of 1. To make the distribution from simulations more directly comparable, as mentioned in Section~\ref{subsec:props}, we assign an observed PDF to each SFR in the simulation by drawing randomly from the observed galaxies with similar masses and SFRs (as the width of the PDF is dependent on these quantities). We then sum the TNG50 PDFs in the same way as the observed ones. In other words we add observational uncertainties to the simulated SFRs so that the scatter is directly comparable. 

Figure \ref{fig:delmsdist} (top row) shows this comparison -- the distribution of simulated and observed galaxies around the main sequence in bins of stellar mass. We remind the reader that observed and simulated galaxies with very low SFRs ($ t_{\rm double} < 20 t_{\rm Hubble}(z) $) are not considered in this analysis. At all masses, the TNG main sequence is narrower than the observations. That is, there is less scatter in the SFRs of the simulated galaxies than there is amongst the observed galaxies.
We note that we use the instantaneous SFR in the simulations and the SFR averaged over 30Myr in observations. The scatter  of  the main sequence  measured from instantaneous SFRs  will be larger than those averaged over  longer  timescales \citep[e.g.][]{Caplar:19,Donnari:19,Tacchella:20} so likely  the difference in the scatter  is even larger than we  see  here. This is less dramatic below log(M$_*$) = 10 and more dramatic above. We quantify this difference in width by computing the width of the region that contains 68\% of the distribution. These values are listed in table \ref{tab:deltams_dist_comp}. For $9<\log({\rm M_*})<10$, we find the simulations are 80-90\% the width of observations. For $10<\log({\rm M_*})<10.5$, the difference grows to 70\%; for $10.5<\log({\rm M_*})<11.$ to 60\%. 

\begin{table}
	\centering
	\caption{Scatter in the star forming main sequence in 3D-HST/\prospector versus TNG50. This is measured in bins of stellar mass with a width 0.5dex including observational uncertainties on both the observations and simulations. (See \S\ref{subsec:sfmswidth}  for more details.)}
	\label{tab:deltams_dist_comp}
	\begin{tabular}{lcccc} 
		\hline
		 mass bins & 3D-HST/\prospector\ & TNG50 & ratio\\
		 \hline
		 $9<\log({\rm M_*})<9.5$ & 0.41 & 0.33 & 0.81\\
		$9.5<\log({\rm M_*})<10$ & 0.38 & 0.33 & 0.88 \\
		$10<\log({\rm M_*})<10.5$ & 0.45 & 0.32 & 0.72\\
		$10.5<\log({\rm M_*})<11$ & 0.57 & 0.33 & 0.58 \\
		\hline
		\hline
	\end{tabular}
\end{table}

The distribution is more skewed toward low SFRs in observations. While in TNG50, the distributions are self-similar at all masses, in observations they become more skewed toward high masses. We quantify this by measuring the skew of the distributions of the simulated vs. observed galaxies based on the ridgeline of the distribution instead of the mean as in the standard definition. At $10.5<\log({\rm M_*/M_{\odot}})<11$, the observed distribution of SFR has a skew of -2.3 while TNG50 has -1.5. The relative lack of low SFR galaxies in TNG50 is likely due to the prescriptions for AGN feedback in the simulation. Although we  note from the observational side that SFRs of low-sSFR galaxies are the most model-dependent. On the simulation side, kinetic radio-mode AGN feedback is designed to very efficiently shut down star formation, while the thermal quasar-mode is comparably inefficient \citep{Weinberger:18}. Within the model, every black hole is in one of these two modes, with low-mass, rapidly accreting black holes (living in low-mass or high redshift galaxies) being in the thermal mode. Once the accretion rate (relative to the Eddington accretion limit) drops below a black hole mass dependent factor, the feedback mode switches to a kinetic mode, leading to an overly sharp decline, almost a jump, in sSFR as a function of black hole mass as well as stellar mass and other properties of the simulated $z\sim0$ galaxy population \citep{Terrazas:20, Habouzit:19, Li:20, Habouzit:21}. We speculate that, similarly, $z\sim1$ galaxies in TNG50 quickly quench whereas in the real  universe massive galaxies seem more likely to tarry below the main sequence before becoming fully quenched. 

Furthermore, at \m>$10^{10}$\msun\ above the main sequence an insufficient number of starbursts as compared to the real Universe  was noted in Illustris \citep{Sparre:15}. We quantify this by comparing the fraction of star formation that occurs more than 2.5$\sigma$ above the main sequence in 3D-HST and TNG50 \cite[as in][]{Rodighiero:11,Sparre:15}. We calculate this fraction based on both ridgelines of the distributions. Our definition is shown in Fig.~\ref{fig:delmsdist} (bottom row). At high masses, TNG50 has a ridgeline which is $\sim0.15$ dex lower than observations despite the medians being the same. We also use the scatter as a function of mass from TNG50 to compute this for both observations and TNG50 because the scatter is significantly smaller in TNG50 than observations at high masses. 
For mass bins [9,9.5],[9.5,10],[10,10.5],[10.5,11] in observations we find the following fractions of star formation occurring in starbursts [12\%,10\%,6\%,2\%] and for TNG50 we find [15\%,13\%,5\%,3\%]. After accounting for the difference between the ridgeline and median of the distribution of SFRs, using the same value for scatter, and accounting for errors on the observed SFRs, the fraction of star formation occurring in ``starbursts" is very similar in observations from 3DHST/Prospector and TNG50. The primary issue with the TNG50 SFRs is the shape of the distribution from the ridgeline to low SFRs. 


\section{sSFR profiles in TNG50 vs. 3D-HST}\label{sect:compprof}

Here we compare the average radial profiles of sSFR in galaxies on, above, and below the SFMS in observations from the 3D-HST survey at $z\sim1$ and the TNG50 magnetohydrodynamical cosmological simulations. The derivation of the profiles is described in \cref{subsec:dataprof} for the observations and \cref{sect:sims} for the simulations. 

The sSFR profiles are a powerful diagnostic to understand where the galaxies are growing. A flat sSFR profiles indicates that the stellar mass doubles at all radii with the same pace, implying a self-similar growth of the stellar mass density profile. An increasing sSFR toward the outskirts implies that the galaxy grows stellar mass faster in the outskirts than in the center (galaxy grows in size), while a decreasing sSFR toward the outskirts is consistent with a galaxy that decreases its size.

\subsection{Inside-out Quenching}
\label{subsec:insideoutquenching}

\begin{figure*}
    \centering
    \includegraphics[width=\textwidth]{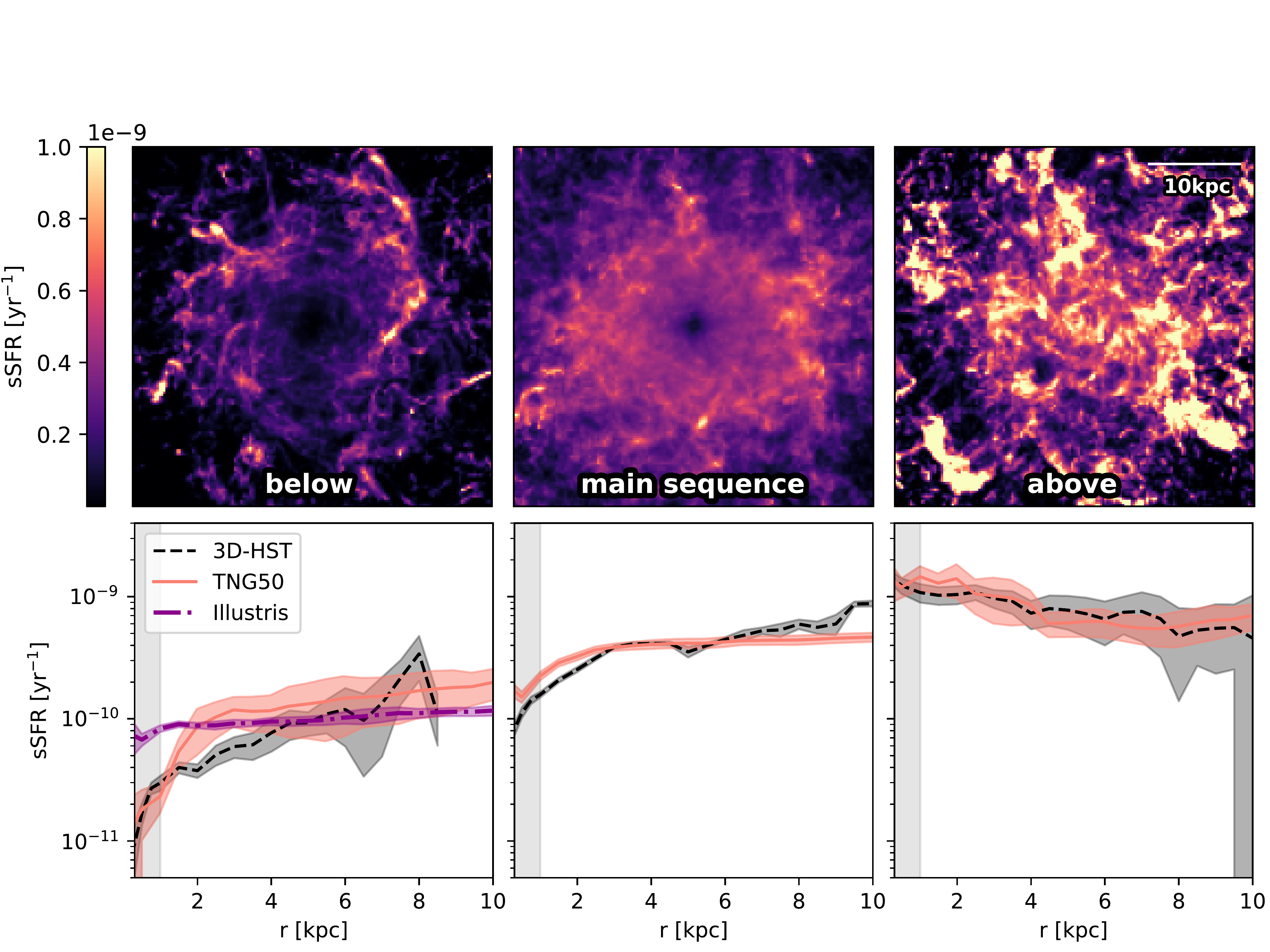}
    \caption{Top row: stacks of sSFR in TNG50 at random orientations. Bottom: specific star-formation rate (sSFR) radial profiles of massive galaxies, with $10^{10.5} < M_\star / \rm{M}_\odot < 10^{11}$ at $z\sim1$. We contrast profiles inferred from observations with 3D-HST (dashed cuves) against the outcome of the TNG50 hydrodynamical simulation (solid curves), as a function of offset from the star-forming main sequence: galaxies which reside below (left), on (center), and above (right). In all cases the TNG50 simulation broadly reproduces both the normalization and shape of the observed SFR radial profiles. A key result of this work is that quenching galaxies (left) exhibit a clear central SFR suppression in the data as well as in TNG50. This supports the scenario of inside-out quenching, which in TNG50 arises due to a central, short time-scale, ejective supermassive black hole feedback mechanism at low accretion rates. This is not the case with the jet-inflated bubble black hole feedback model in Illustris as shown by the dash-dot purple line. The grey shaded region is inside  the observed PSF.}
    \label{fig:ssfrprofs_highmass}
\end{figure*}

A key result of this paper is that star formation is quenched from the inside-out, which in the simulations is caused directly by AGN feedback. Figure \ref{fig:ssfrprofs_highmass} shows that below the main sequence at 10.5$<$log(\m)$<$11, the sSFR profiles are strongly centrally suppressed in both TNG50 and in observations \citep[e.g.][]{Nelson:16b,Tacchella:18a,Ellison:18,Belfiore:18,Morselli:19}. In TNG50, this centrally suppressed star formation is a key signature of locally acting AGN feedback \citep[see also][]{DNelson:19}. 

\begin{figure*}
    \centering
    \includegraphics[width=\textwidth]{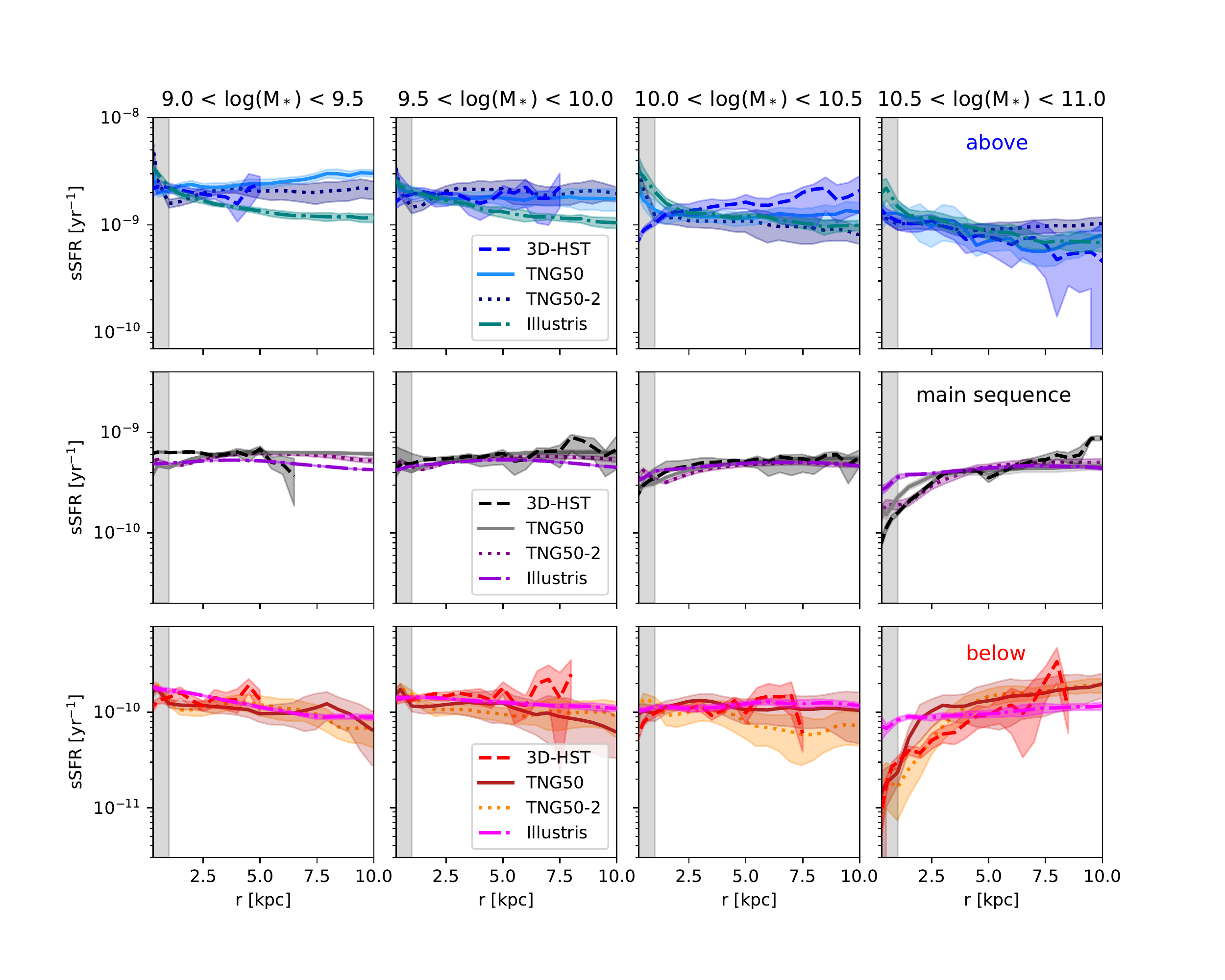}
    \caption{sSFR profiles of $z\sim1$ galaxies across the star forming main sequence -- comparison between observations and TNG50, the original Illustris simulation, and a lower resolution version of TNG50 with resolution more similar to that of Illustris (TNG50-2). Profiles are cut off when their signal-to-noise ratio falls below 1. We find that TNG50 is more consistent with observations than the original Illustris simulation and that this is not primarily due  to resolution effects. The grey shaded region is inside  the observed PSF.}
    \label{fig:profs_comp_resog}
\end{figure*}

This signature is not seen in the original Illustris simulation, where AGN feedback acts non-locally. In Fig.\,\ref{fig:profs_comp_resog} we also disentangle the impact of resolution, comparing TNG50-1 to TNG50-2, the analogous simulation run with eight times lower mass resolution (two times lower spatial resolution). As shown through the comparison to the lower resolution version of TNG50, this is not a resolution effect but due to the physics in the simulation. In Illustris, bubbles are blown at galactocentric distances of 50-100 kpc and consequently have a hard time propagating back into the denser gas to affect the center of the galaxy. Hence the sSFR profiles in Illustris are not centrally suppressed. In TNG50 on the other hand, both kinetic and thermal feedback are done on the gas immediately surrounding the black hole, suppressing star formation from the inside-out. 

In quantitative detail there remain small differences between the observed and TNG50 sSFR profiles at high masses (i.e. log \m > 10.5) below the main sequence. While the sSFR profiles agree at the centers, for $2<r<4$\,kpc TNG50 is about a factor of two higher than observations. This implies that the central suppression of SFR does not extend to sufficiently large radii as seen in the data.
This could be related to the modeling of the interstellar medium in this simulation: in particular, there is no explicit multi-phase medium with cold clouds embedded in a hot, volume filling component, but cells have a single, volume averaged density value and a pressure according to an effective equation of state \citep{Springel:03}. This means that AGN driven winds that interact with this medium impact the entire mass budget, while a situation where the wind propagates in low density channels while cold clouds continue forming stars \citep[e.g.][]{Dugan:17} is not possible within the IllustrisTNG model. It is possible that the effect of AGN winds would differ with a more realistically modeled ISM -- a scenario testable with future simulations. 

In general the TNG AGN feedback model produces sSFR profiles which are in better agreement with observations than the original Illustris simulation. 
The TNG black hole feedback model introduces powerful  kicks when a black hole reaches a certain mass \citep{Weinberger:17, DNelson:19}. These kicks evacuate gas from the very center of the galaxy \citep{Zinger:20}, introducing enough feedback energy to gravitationally unbind gas from the galaxy. 
As described in \cite{Terrazas:20}, these galaxies are likely in the process of unbinding their gas starting from the very central regions and eventually expanding its effect to larger radii. 
Notably the original Illustris simulation, with its rather different physical mechanism for AGN feedback at low accretion rates, based on jet-inflated bubbles heating the ICM at distances of tens of kpc or more from the galaxy, does not reproduce the central SFR profile suppression seen in data. This different manifestation between the two feedback models is clearly constrained by the observations. In summary, our findings support two key ideas: (i) in reality, massive galaxies quench from the inside-out possibly due to locally acting AGN feedback, while (ii) in the TNG simulations, the details of how supermassive black hole feedback are implemented and, in particular, how this feedback energy physically affects, heats, and redistributes gas appears to zeroth order consistent with constraints from the observed star formation rate radial profiles on scales of a kiloparsec.

\subsection{Flat sSFR profiles across the star-forming main sequence} \label{subsec:profileAgreement}

\begin{figure*}
    \centering
    \includegraphics[width=\textwidth]{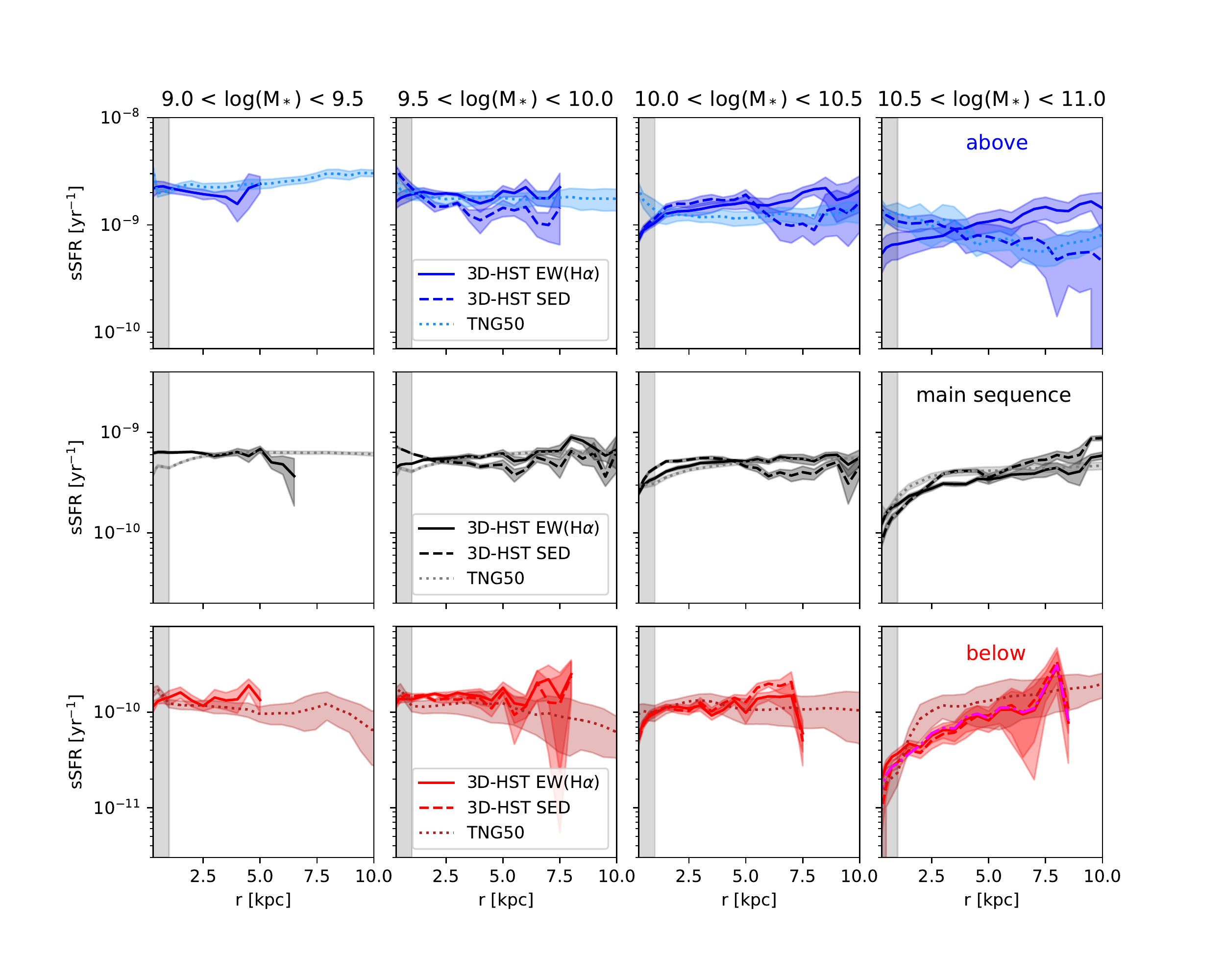}
    \caption{The average radial sSFR profiles of galaxies across the star forming main sequence are very similar between TNG50 and observations at $0.75<z<1.5$. The top row is  above the main sequence, middle is on, bottom is  below. The magenta in the bottom right corresponds to the AGN correction described in \S\ref{subsec:dataprof}, note it makes little difference. The grey shaded region is inside  the observed PSF.}
    \label{fig:profs_comp}
\end{figure*}

Average specific star formation rate (sSFR) profiles of galaxies on, above, and below the star forming main sequence in observations and simulations are shown in Fig.\,\ref{fig:profs_comp}. The main takeaway is that the sSFR profiles across the main sequence in TNG50 are remarkably similar to those in observations. With few exceptions, at all masses and radii the observed and simulated sSFR profiles lie within 0.3 dex (a factor of two) of each other. 

This agreement is surprising; it did not have to turn out this way. The consistency shows that the distribution of dense gas and the conversion of gas into stars are roughly correct in the simulation, at least relative to the existing stellar mass. This means that the physical TNG50 model governing how galaxies grow in size and build their structures across the SFMS yield high fidelity predictions. The distribution of cold gas is set by the spatially dependent interplay between gas inflows, outflows, and star formation.
The accretion of gas onto the galaxies is driven by gravity (a model about which there is less uncertainty than the others) and suppressed by feedback. TNG50 uses the \cite{Springel:03} model for star formation. In this model  gas above a certain density threshold is converted to stars stochastically. While this model is too simple on small scales \citep[e.g.][]{Semenov:19}, it appears that on kpc scales this model produces results that are consistent with observations. 

Feedback has significant effects in all parts of the baryon cycle: it affects inflow rates and geometries \citep[e.g.][]{DNelson:15} and it determines the distribution of cold gas and hence where the galaxy can form stars. In TNG50 outflows driven by supernova feedback are launched from star forming gas with their energy given by the star formation rate. This result means that at $0.75<z<1.5$ and 9<log(\m)<10.5, the way outflows are implemented in TNG50 produces results that are on population and azimuthal average, consistent with observations on, above, and below the star forming main sequence. TNG50's combination of outflows and conversion of gas into stars produces galaxies that have a radial structure of new star formation over past star formation that is consistent with the real Universe.

Above the main sequence the sSFR profiles from TNG50 and  3D-HST are fairly flat. Star formation is not primarily enhanced in the center meaning that it is not primarily driven by central starbursts. In this regime, the match with observations improved from Illustris to TNG. In Illustris the profiles have somewhat of a negative gradient while in TNG50 (and 3D-HST observations) they do not. This is not  primarily a resolution effect as the profiles in TNG-LowRes are fairly flat like those in TNG50. Instead this is likely a physical effect owing to the implementation of supernova feedback. As shown in obsevations as well as in TNG50  \citep{Forster-Schreiber:19,DNelson:19}, star formation driven winds are strongest above the SFMS, at least at $z\sim1$. The implementation of these winds changed from Illustris to TNG. As shown in \cite{Hemler:20}, this affects the metallicity gradients in galaxies. Here we see that it also affects the shape of the sSFR profiles of galaxies above the main sequence. In TNG50 wind energy has an additional scaling with the metallicity \citep{Pillepich:18a}. These changes produce flatter sSFR profiles above the main sequence, more in line with observations from  the 3D-HST survey at $z\sim1$.

What do the shapes of the sSFR profiles mean for how galaxies build structurally? Across the main sequence at all masses and star formation rates, the sSFR profiles on average are fairly flat, meaning that galaxies grow largely self-similarly on average \citep{Nelson:19}. This is consistent with the fact that the size-mass relation for star-forming galaxies has a shallow slope \citep[e.g.][]{vanDokkum:13,vanDokkum:15,Patel:13,Suess:19,Mosleh:20}. Star formation adds stars to galaxies with close to the same distribution as the existing stars so the structure of galaxies as a population as a function of mass changes fairly slowly. This is not necessarily true of individual galaxies and in fact the purpose of this detailed comparison between observations and simulations is in service of the ability to use these simulations to track individual galaxies through time to see what drives their evolution through the SFR-\m\ plane. 

\begin{figure*}
    \centering
    \includegraphics[width=\textwidth]{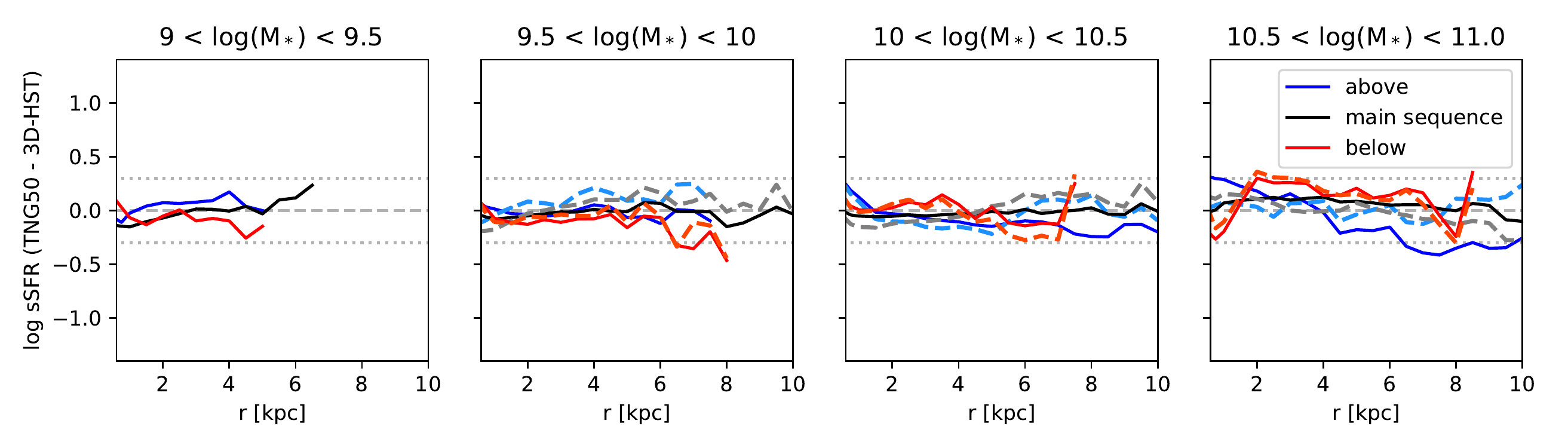}
    \caption{Difference between the sSFR profiles in 3D-HST and TNG50. As shown by the dotted grey lines, the profiles are nearly always within $\pm0.3$dex (a factor of two) of each other.} 
    \label{fig:ssfr_diff}
\end{figure*}


\section{Summary}\label{sect:summary}

In this paper we have compared the integrated and kpc-resolved star forming main sequence in the TNG50 magnetohydrodynamical cosmological simulation and observations from the 3D-HST survey. TNG50 is the highest resolution iteration of the IllustrisTNG project, resolving 2100 galaxies with \m$>10^9$ \msun\ at a spatial resolution of $\sim100$ pc at $z\sim1$. The 3D-HST program is a 248 orbit near-infrared spatially resolved spectroscopic survey with the Hubble Space Telescope that provides maps of the specific star formation rate in thousands of galaxies at $z\sim1$. These are complemented by a new analysis of the integrated photometry of these galaxies with the \prospector\  Bayesian inference framework, providing improved estimates for stellar masses and SFRs. These simulated and observed datasets are well-matched to determine how well the simulation can be used to understand how galaxies move through the star forming main sequence, what causes star formation to be enhanced and suppressed, and how galaxies evolve structurally during this process. 

We find that the star forming main sequence in TNG50 is consistent to within 0.1 dex of observations from 3D-HST for all masses $10^9<$\m$<10^{11}$\msun\ at $0.75<z<1.5$ derived from \prospector. This is a significantly stronger agreement than previously reported for the TNG simulations in comparison to then-available observationally-inferred results \citep{Donnari:19} and a strong validation of the model in a galaxy integrated population sense (see Fig.\,\ref{fig:sfms_comp}). This is also better than the agreement typically reported in cosmological hydrodynamical simulations \citep{Torrey:14,Sparre:15,Schaye:15,Somerville:15,Dave:16}. 
We find that the previous 0.2-1 dex offset between observations and simulations may be driven by the inference of stellar population parameters from observations rather than necessarily the physical model in simulations, although uncertainties remain to be tested regarding star formation histories and other aspects of the inference of stellar populations. The newly-derived stellar mass estimates are 0.1-0.3\,dex higher and the star formation rates 0.1-1\,dex lower than previous estimates \citep[see][for more details]{Leja:19b}.
While the median SFRs are nearly identical between TNG50 and observations, some discrepancies do arise in the higher order moments of the SFR distribution. The scatter of SFRs around the main sequence in TNG50 is narrower at all masses than in observations. It is also self-similar while the observed SFRs skew towards lower values as mass increases (Fig. \ref{fig:delmsdist}). 

Further, we find surprisingly good agreement between the simulated and observed average sSFR radial profiles of galaxies above, on, and below the star forming main sequence. With a few exceptions, they agree qualitatively and quantitatively. They are within a factor of two at all masses and radii across the main sequence. Qualitatively, in both observations and simulations, across the main sequence, the sSFR profiles are fairly flat, meaning galaxies on average grow self-similarly regardless of where they are in the SFR-M$_*$ plane, which is likely why the size growth of galaxies is so gradual.
This means, importantly that the distribution of gas and its conversion into stars in the simulation are at least roughly correct on kpc scales. 

The agreement between TNG50 and 3D-HST data is particularly interesting below the main sequence at high masses, a region of parameter space that galaxies must necessarily traverse on their journey from star forming to quenched. Here we find that both simulated and observed $z\sim1$ galaxies exhibit depressions in sSFR in the central regions, up to a few kpc wide. The inside-out suppression of star formation in high mass galaxies below the main sequence is similar in both 3D-HST observations and in the TNG50 simulation, a key signature of locally acting AGN feedback. This behavior is not seen in the original Illustris simulation, where AGN feedback  affects gas at large radii rather than acting directly from the innermost regions of galaxies. Taken together, our results provide evidence for AGN feedback as the source of galaxy quenching. 

Looking ahead, because the simulation reasonably reproduces the observations, we should be able to use the simulation to understand how galaxies move through the SFR-\m\ plane and build structurally through star formation. 

\section*{Acknowledgements}
The TNG50 simulation was realized with compute time granted by the Gauss Centre for Supercomputing (GCS) via the Large-Scale Project GCS- DWAR (2016; PIs Nelson/Pillepich); its lower resolution counterparts were carried out on the Draco and Hydra supercomputers at the Max Planck Computing and Data Facility (MPCDF); the original Illustris simulation was performed at the CURIE supercomputer at CEA/France as part of PRACE project RA0844 and at the SuperMUC computer at the Leibniz Computing Centre, Germany, as part of project pr85je. EJN acknowledges support of the National Hubble Fellowship Program through grant number HST-HF2-51416.001-A. ST is supported by the Smithsonian Astrophysical Observatory through the CfA Fellowship. BB acknowledges support of the Simons Foundation Flatiron Institute and is a Packard Fellow. FM acknowledges support through the Program "Rita Levi Montalcini" of the Italian MUR. BAT was supported by the Harvard Future Faculty Leaders Postdoctoral Fellowship. The Cosmic Dawn Center (DAWN) is funded by the Danish National Research Foundation under grant No.\ 140. RKC acknowledges funding from the John Harvard Distinguished Science Fellowship


\bibliographystyle{mnras}
\bibliography{all}


\appendix

\section{Appendix}

\begin{figure*}
    \centering
    \includegraphics[width=\textwidth]{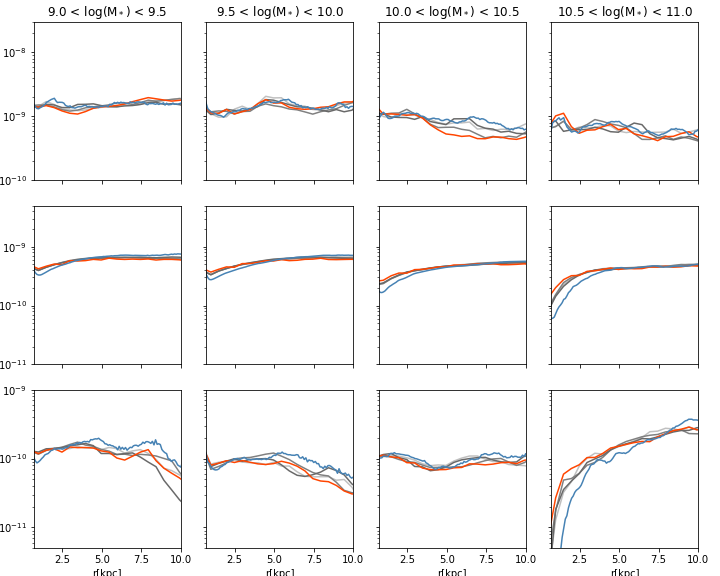}
    \caption{Here we show the difference between sSFR profiles from TNG50 at different orientations. Blue is face on, orange is edge on, grays are xy,xz,and yz projections respectively. The top row is above the main sequence, the middle is on the main sequence, the bottom is below.}
    \label{fig:ssfrprofs_orients}
\end{figure*}

As noted in \S\ref{subsec:tngprofs}, in Fig.~\ref{fig:ssfrprofs_orients} we show the impact of orientation on average sSFR profiles across the SFMS. The primary region of parameter space where this turns out to be relevant is also the most interesting: at high masses below the main sequence. Projection effects result in sSFR profiles that appear less centrally depressed than they are in reality if one could measure them face-on. This is relevant for our interpretation of observations.


\label{lastpage}
\end{document}